\documentclass[journal]{IEEEtran}
\usepackage{graphicx}
\usepackage{amsmath, amsfonts}
\usepackage{url}
\usepackage{hyperref}
\usepackage{cite}
\usepackage{multirow}
\usepackage{booktabs}
\usepackage{subcaption}
\usepackage{paralist}
\usepackage{booktabs}
\usepackage{xcolor}
\usepackage{soul}

\usepackage{pgfplots}
\pgfplotsset{width=8.5cm, compat=1.9}
\pgfplotsset{every tick label/.append style={font=\tiny}}

\graphicspath{{./figs/}}

\begin{document}

\title{Detecting Parkinsonian Tremor from IMU Data Collected 
In-The-Wild using Deep Multiple-Instance Learning}

\author{
  Alexandros Papadopoulos$^1$,~\IEEEmembership{Student~Member,~IEEE,}
  Konstantinos Kyritsis$^1$,~\IEEEmembership{Student~Member,~IEEE,}\\
  Lisa Klingelhoefer$^2$,
  Sevasti Bostanjopoulou$^3$,
  K. Ray Chaudhuri$^4$,
  Anastasios Delopoulos$^1$,~\IEEEmembership{Member,~IEEE,}
  \thanks{$^1$Multimedia Understanding Group, Information Processing Laboratory, Dept. of Electrical and Computer Engineering, Aristotle University of Thessaloniki, Greece
  $^2$Department of Neurology, Technical University of Dresden, Dresden, Germany
  $^3$Third Neurological Clinic, Papanikolaou Hospital, Thessaloniki, Greece
  $^4$International Parkinson Excellence Research Centre, King's College
  Hospital NHS Foundation Trust,
  London, UK
  $^5$ Copyright 2019 IEEE. Personal use of this material is permitted. Permission from IEEE must be obtained for all other uses, in any current or future media, including reprinting/republishing this material for advertising or promotional purposes, creating new collective works, for resale or redistribution to servers or lists, or reuse of any copyrighted component of this work in other works}%
}

\maketitle

\begin{abstract}
  Parkinson's Disease (PD) is a slowly evolving neurological disease
  that affects about $1\%$ of the population above 60 years old, causing
  symptoms that are subtle at first, but
  whose intensity increases as the disease progresses.
  Automated detection of these symptoms could offer clues as to the early
  onset of the disease, thus improving the expected clinical outcomes of
  the patients via appropriately targeted interventions. 
  This potential has led many researchers to develop 
  methods that use widely available sensors to 
  measure and quantify the presence of PD symptoms 
  such as tremor, rigidity and braykinesia.
  However, most of these approaches operate under
  controlled settings, such as in lab or at home, thus limiting their 
  applicability under free-living conditions. 
  In this work, we present a method for automatically identifying tremorous episodes
  related to PD, based on IMU signals captured via a smartphone
  device. We propose a \emph{Multiple-Instance Learning} approach, 
  wherein a subject is represented as an unordered bag of
  accelerometer signal segments and a single, expert-provided, tremor
  annotation. 
  Our method combines deep feature learning with a learnable pooling
  stage that is able to identify key instances within the subject bag,
  while still being trainable end-to-end.  
  We validate our algorithm on a newly introduced dataset of $45$ subjects,
  containing accelerometer signals collected entirely in-the-wild. 
  The good classification performance obtained in the conducted experiments
  suggests that the proposed method can efficiently navigate the
  noisy environment of in-the-wild recordings.
\end{abstract}

\IEEEpeerreviewmaketitle

\section{Introduction}
%
%
%
%
\IEEEPARstart{P}{arkinson's} Disease is a long-term neurodegenerative condition 
that targets the central nervous system. Its symptomatology includes 
motor symptoms, such as tremor, bradykinesia, rigidity and hypomimia, as
well as
symptoms of non-motor nature like constipation and insomnia. In particular, tremor,
bradykinesia and rigidity have been characterized as cardinal symptoms, 
in the sense that regardless of the symptom variability 
across different cases, the co-occurrence of at
least two of these symptoms is a good indicator
of the disease\cite{Jankovic368}.

Despite being incurable, early diagnosis of PD holds much clinical benefit,
since symptoms in earlier stages can be managed more efficiently through
appropriately targeted interventions\cite{pagan2012improving}. 
However, the initial symptom onset can be so subtle
that it goes unnoticed by the subjects until the disease has
already progressed.
Therefore, development of automated tools and methods that can observe and quantify the
severity of PD symptoms outside laboratory conditions is a 
very useful research direction.

Following this line of research, many methods to 
automatically detect PD symptoms using data
captured from a variety of sensors
have been proposed. For instance,\cite{orozco2015voiced} and
\cite{6126094} use microphone-recorded speech signals to
estimate the severity of speech impairment that exhibits in some PD
patients using pre-computed features and Random Forest classifiers, 
while\cite{deep_speech} proposes a CNN-based deep learning 
architecture that uses different CNN modules on different sets of features
and then merges the learned representations.
In a similar vein, the work of\cite{giancardo2016computer} uses a physical
keyboard as a capturing device and proposes a method to transform the
recorded sequence of key taps
to a PD motor index. This approach is further refined by\cite{iakov} who
use the virtual keyboard from a smartphone
and correlate the resulting index with the severity of bradykinesia and
rigidity. Hand dexterity is also used as a means of PD diagnosis by the
work of \cite{deep_handwriting}, who 
capture handwritten dynamics via a smart pen sensor and use it to train 
a CNN model to differentiate healthy from PD subjects.
Many works employ \emph{Inertial Measurement Unit (IMU)}
sensors for data capturing. 
For example, \cite{imu_gait} examines the feasibility 
of using body-worn IMU sensors for performing gait analysis,
with the ultimate goal of inferring whether the wearer is a PD patient.
In a similar spirit, \cite{emd} \cite{s100302129}
and\cite{dai2015quantitative} make use of stand-alone accelerometer and gyroscope
sensors to quantify tremor severity, while\cite{tremor_phone}
explores the potential of using the IMU sensors embedded in smartphones as a
viable means of monitoring and detecting tremor.

The idea of using sensors embedded in commercial off-the-self devices, such
as smartphones and smartwatches is widely explored by \emph{i-PROGNOSIS}
\cite{lisa_klingelhoefer_2017_1199554}, a European Horizon 2020 research
programme that adopts a holistic approach towards early Parkinson's
diagnosis. In particular, it utilizes 
a multi-modal approach,
where data are collected unobtrusively and in-the-wild, through smartphone
embedded sensors including IMU,
microphone and virtual keyboard, and consequently 
mapped to PD symptom indicators using
machine learning techniques.
Thus, rather than inferring the presence of PD from single symptom
clues, i-PROGNOSIS examines the user's status with respect to multiple
symptoms, as well as, their longitudinal evolution, in order to detect
the disease onset.

In this paper, we focus on the problem of automatically detecting PD tremor
from IMU data collected in-the-wild via a smartphone. Tremor detection
in-the-wild, has received little attention from the research community,
owing to the inherent difficulties of obtaining and properly annotating
datasets in that setting. To this end, we introduce a
new dataset of accelerometer recordings from both PD patients and healthy
users, captured outside laboratory conditions and without any form of
supervision or guidance, via a smartphone application developed in the
context of\cite{lisa_klingelhoefer_2017_1199554}. 

Unlike other symptoms that are present all the time, tremor is of
intermittent nature, that is, it may come and go unpredictably or, more
formally, exhibit on and off periods. This problem is usually circumvented
by recording the data acquisition sessions with a camera. The video is then
used by medical experts to provide a fine-grained annotation of the on and
off periods. However, this solution 
does not apply to data obtained unobtrusively in-the-wild, where observing
the subjects throughout their daily lives is impossible. Hence, in that
setting we must make do with the coarse label provided by the medical
experts during the examination of the subject.
Unfortunately, we
cannot associate each IMU recording provided by a subject with this coarse tremor
ground truth, as that would introduce
severe label noise to our models that would prove devastating during training.
Therefore, to mitigate this issue, 
we propose to address the problem at hand as a case
of \emph{Multiple-Instance Learning (MIL)}.

Multiple-Instance Learning \cite{dietterich} is a supervised learning scenario, where multiple
instances or data points are grouped together to form sets, commonly
referred to in MIL terminology
as \emph{bags} \cite{mil_survey}. Each bag is associated with a single label 
that depends on the labels of the instances it contains according to some
assumption. For example, the standard MIL
assumption that we adopt in this paper, 
dictates that for a binary classification problem, 
a bag is considered positive if it contains at least one positive instance
and negative if contains only negative instances.
In our case, however, we are afforded with only an approximation of the
true bag labels that is provided by an external 
oracle (be it the medical expert or the subject itself), 
who operates without any knowledge of the individual instance labels.
For a detailed review of the different MIL assumptions, as well as the most
popular algorithms and applications, 
we refer the reader to the survey of
\cite{mil_survey}.

We can use the formalism of MIL to tackle the problem of tremor
detection in-the-wild. The MIL framework naturally deals with the lack of
fine-grained annotations (instance labels) and thus we can use it to
adequately handle
the particular nature of tremor: each subject can be represented as a
bag of accelerometer signal segments and a tremor annotation that describes
the whole bag. We propose a method to efficiently model the
probability that a subject has tremor given their bag of acceleration
segments, based on a recently proposed MIL method\cite{pmlr-v80-ilse18a},
that makes use of the attention mechanism\cite{NIPS2017_7181} 
to identify key instances within a bag. 
The proposed method builds upon a 
previous work of ours \cite{papadopoulos2019mil},
that showed very promising early results. In this paper, we extend that
work by:
\begin{enumerate}[i)]
  \item Enriching the original dataset with $8$ additional subjects for a
    total of $45$ subjects.
  \item Introducing an alternative approach that operates on the raw
    accelerometer values and outperforms the originally proposed
    spectrogram-based method.
  \item Performing extensive experiments to confirm the potential of our
    method.
  \item Publishing the dataset as open access data 
    (available in \url{https://zenodo.org/record/3519213}.)
\end{enumerate}

The rest of this paper is organized as follows. Section \ref{soa} discusses
the most relevant work regarding in-the-wild tremor detection from IMU
sensors. Section \ref{method} presents the proposed tremor detection
methodology. Section \ref{dataset} describes the dataset used in our
experiments. Section \ref{experiments} discusses the experimental setup and
presents the results of our method along with those of other popular alternatives.
Section \ref{discussion} contains a critical discussion of the proposed
method.
Finally, the paper concludes with section \ref{conclusion}.

\section{Related work}\label{soa}
Parkinsonian tremor is an involuntary muscle contraction that typically
exhibits at a frequency of $3\hbox{-}7$ Hz. Its particular nature makes the use of inertial
sensors, such as the accelerometers and gyroscopes embedded in modern phones, 
particularly appealing for
developing data-driven tremor detection methods. Although many works
have explored the use of such sensors for tremor detection, the research
community has devoted little attention to developing algorithms that
operate well outside laboratory conditions. Most approaches, such as 
\cite{arora2014high}, \cite{arora2015detecting}, are restricted to the home
environment, while the evaluation of the symptom
severity takes place at pre-defined moments during the day via self-administered
tests, rather than occurring unobtrusively.

One of the works most related to ours is that of \cite{das2012detecting},
in which the authors actually proposed a multiple-instance learning scheme to
detect PD motor symptoms based on $5$ body-worn accelerometers. Two
subjects were tasked with wearing the sensors constantly for
a period of four days while completing their regular daily living
activities. The subjects were also asked to keep a journal of their
medication intake and approximate time intervals on which symptoms occurred
After data collection, the accelerometer stream was segmented to windows of $6$
seconds length and $1$ second overlap, and several hand-crafted features
were extracted from each window. All the window feature vectors
pertaining to a specific time interval (of typical duration of $20-40$
minutes) were used to create a bag whose label was defined by 
the symptom occurrence entry for that interval in the subject's journal.
The \emph{Axis-Parallel Hyper-Rectangle (APR)} \cite{mil} 
algorithm was then trained using the acquired labeled bags from the first $8$
hours of the first day of monitoring. Evaluation on the data recorded
during the other days showed promising accuracy. However, the authors do not
provide details about the class imbalance 
or the metrics affected by it (recall, specificity).

The more recent work of \cite{zhang2017weakly}, extended the previous
approach by using a dataset of 5 PD patients all of which
exhibited tremor. Data collection took place under laboratory
conditions that resembled a home environment. In that setting, the subjects
were free to perform any from a list of given activities, such as walking,
writing or playing chess, while multiple cameras were used to provide detailed
annotation regarding the tremor intensity at each moment. Standard
hand-crafted features were extracted over $2$-second windows with
$1$ second overlap. A window was considered tremorous if at least half
contained a tremor episode as identified by the video. Consecutive window
feature vectors were then pooled together into larger segments of length varying
from $30$ seconds to $10$ minutes, to form a bag. Each such bag was
labeled using two different methods: i) a bag was considered positive if
it contained at least one positive window i.e. labeling according the 
standard multiple-instance learning label assumption ii) a stratified method that
considered the approximate percentage of tremor within a segment, quantized
in $4$ levels corresponding to $0$-$24\%$, $25-49\%$, $50-74\%$ and
$75-100\%$. The authors performed leave one subject out experiments with classical
multiple-instance learning algorithms to show that performance rapidly
deteriorates for the standard labeling approach as the segment length
grows. Finally, they proposed a simple modification, applicable
to all multiple-instance algorithms,
that takes advantage of stratified labels so as to avoid the deterioration
in performance caused by the decreasing label precision owing to the
increase of the segment length.

While similar in spirit, our work differs from the previous approaches
significantly. First of all, we use a large dataset of $45$ subjects 
that contains both PD patients, with and without tremor, 
as well as completely healthy individuals. 
In addition, our dataset was collected unobtrusively and under 
completely unscripted and in-the-wild 
conditions. This means that the noise levels are expected to be 
significantly higher compared to data collected in lab or at home.
Finally, contrary to other works, where fine-grained labels are available, 
we only have access to coarse subject-level labels, provided by
experts. Therefore, the main contributions of this paper are:
\begin{enumerate}
    \item A method for binary detection of Parkinsonian tremor from IMU
      data collected in-the-wild, that employs deep neural networks for
      feature extraction as well as a learnable pooling stage that can
      produce robust subject embeddings, leading to high classification
      performance.
    \item A new, challenging dataset of IMU recordings,
      collected by a population of $45$ PD and healthy subjects 
      under completely unscripted and real-life conditions. The dataset
      is available online in \url{https://zenodo.org/record/3519213}.
\end{enumerate}

\begin{figure*}[!ht]
  \centering
  \includegraphics[width=1\linewidth]{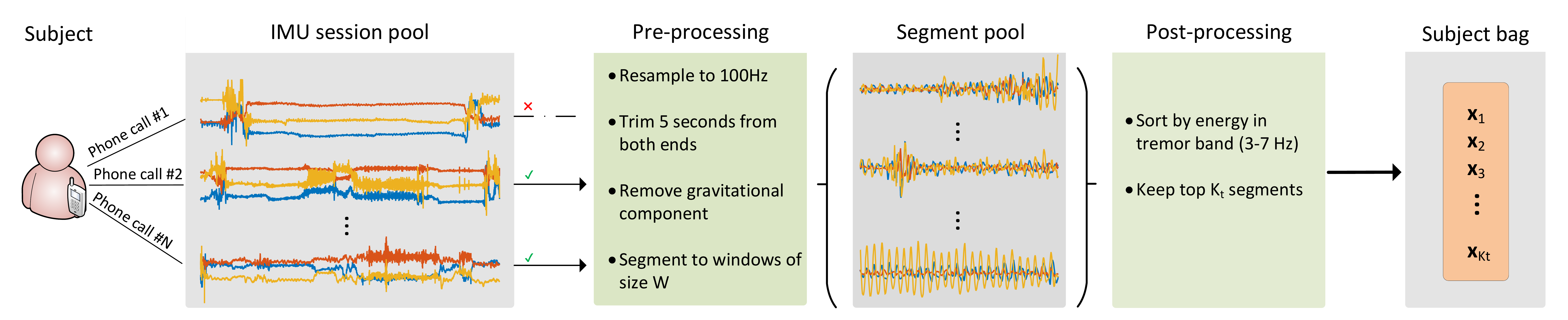}
  \caption{Overview of the bag creation process. Each subject is assigned a
    bag of accelerometer signal segments captured during phone
    calls, and a single tremor label provided by experts.
  }
\label{bag_creation}
\end{figure*}

\section{Multiple-Instance Tremor Detection}\label{method}
Supervised learning is the learning situation where we are provided with a set of
instances $\mathbf{x}_i \in \mathbb{R}^N$ and their corresponding labels
$y_i \in \mathcal{Y}$, and the goal is to infer a general mapping $f:
\mathbb{R}^n \to \mathcal{Y}$. 
The label space $\mathcal{Y}$ can be
either $\mathbb{R}$ (for regression problems) or $\mathbb{Z}$ (for
classification problems). 
The mapping $f$ is usually chosen from a wide class of functions
$\hat{f}(x; \theta)$ through an optimization procedure that minimizes some
suitable cost function on the given set of instances, in order to find
good values for the parameters $\theta$. In the Multiple-Instance Learning
setting, instead of individual instances, we are presented with unordered
sets of instances $X_j = \{ \mathbf{x}_{j1},
\mathbf{x}_{j2}, \dots, \mathbf{x}_{jK_j} \} $, called \emph{bags}, and their corresponding bag-level
labels $y_j$. The goal in this scenario is to infer a mapping $f:
2^{\mathbb{R}^N} \to \mathcal{Y}$, where $2^{\mathbb{R}^N}$ denotes the
power set of $\mathbb{R}^N$, that is to perform classification or
regression on the bag level. In the rest of this work, we will assume that $\mathcal{Y}
= \{ 0, 1 \}$, since our focus is the case of binary classification of
non-tremor (class $0$) vs tremor (class $1$). Hence, in our case the
learned function $f$ will map a given bag to the probability that it belongs 
to the positive class:
$f(X;\theta) = p\textsubscript{model}(y=1|X) = 1 - p\textsubscript{model}(y=0|X)$. 

Since we are dealing with unordered sets it is evident that the learned
function $f$
must be permutation invariant to the elements in a bag. 
The recent work of \cite{NIPS2017_6931} provides a general strategy for
modeling almost any permutation invariant function over a set $X$, as a sum
decomposition of the form 
$f(X) = \rho \left( \mathop{\sum}_{\mathbf{x} \in X}\phi\left(\mathbf{x}
\right) \right)$, given suitable transformations $\phi$, $\rho$.
Thus, we can model the bag label probability as:
\begin{equation}
  p\textsubscript{model}(y=1|X) = \rho \left( \mathop{\sigma}_{\mathbf{x} \in X}\left(\phi\left(\mathbf{x}\right)
  \right) \right)
  \label{score_function}
\end{equation}
where in our case:
\begin{enumerate}[i)]
\item $\phi: \mathbb{R}^N \rightarrow \mathbb{R}^M$ independently maps
  each instance $\mathbf{x}_i$ of $X$
  to a low-dimensional embedding of size $M$.
\item $\sigma: 2^{\mathbb{R}^M} \rightarrow \mathbb{R}^M$ is a
  weighted average of the instance embeddings 
  that produces a fixed-length bag representation.
\item $\rho: \mathbb{R}^M \rightarrow \left[0, 1 \right]$ transforms the
  pooled bag representation to the final bag label probability.
\end{enumerate}

Equation \ref{score_function} allows for increasingly flexible model
designs.
In particular, the function $\phi$ can be seen as performing feature extraction on each
instance, while the function $\rho$ is a classifier entity that outputs the
final class estimation. In this work, we use neural networks to parameterize these two
functions and adopt a learnable pooling
scheme to compute the 
weight of each instance in the pooling function $\sigma$.
More specifically, let $H = \{\mathbf{h}_1, \mathbf{h}_2, \dots,
\mathbf{h}_K \} = \{\mathop{\phi}(\mathbf{x}_1),
\mathop{\phi}(\mathbf{x}_2), \dots, \mathop{\phi}(\mathbf{x}_K)\}$ be a bag
of $K$ embeddings that results from the elementwise application of the
embedding function $\phi$ to the initial bag $X$. According to
equation \ref{score_function}, the function $\sigma$ is defined as:

\begin{equation}
\mathbf{z} = \sigma(H) = \sum_{k=1}^{K} a_k \mathbf{h}_k
  \label{pooling}
\end{equation}
We can integrate learning of the weights $a_k$ in the training procedure, by
using a version of the attention mechanism \cite{NIPS2017_7181}, modified
appropriately in \cite{pmlr-v80-ilse18a} to work for MIL tasks, that takes
the form: 
\begin{equation}
  a_k = \frac{\exp{(\mathbf{w}^T \tanh{(\mathbf{V}\mathbf{h}_k^T}) } )}
  {\sum_{k=1}^{K}\exp{(\mathbf{w}^T \tanh{(\mathbf{V}\mathbf{h}_k^T}) })}
  \label{attention}
\end{equation}

Computation of the quantities $a_k$ is a highly non-linear
procedure, since the value of each $a_k$ depends both on the value of the relevant
embedding $\mathbf{h}_k$, as well as the learnable parameters $\mathbf{w}
\in \mathbb{R}^{L \times 1}$ and $\mathbf{V} \in \mathbb{R}^{L \times M}$,
via a composition of non-linear functions. As we will see in section \ref{experiments}, 
the construction of Equation \ref{attention} is capable of identifying 
and assigning large weights to key instances within a bag, allowing the
final classifier stage (the function $\rho$) to perform efficient bag
classification. This allows us to identify
which instances contributed the most to the model's decision, thus adding a
degree of interpretability to the method. 

We propose to apply the attention-based multiple-instance framework
described above to the problem of detecting Parkinsonian tremor from
data collected in-the-wild.
The rest of this section
outlines the process of creating a bag for each subject using their contributed IMU
data and the training procedure of our tremor detection model.

\subsection{Bag creation}\label{bag_creation_section}
During data collection, each subject contributes one tri-axial accelerometer session 
for every phone call they make during their participation period. 
Each phone call may be of variable duration,
which means that the recorded signals will be of variable length.
However, the bag instances are assumed to be of a specific dimensionality
($\mathbf{x}_i \in \mathbb{R}^N$).
To circumvent this issue, we resort to windowing each session into
non-overlapping segments of length $W$ samples. In doing so, we choose not to
model possible intra-session dependencies between neighboring segments.
However, based on our experimental results, this does not affect the
method's performance.

Let $S_i = \{\mathbf{w}_{1}, \mathbf{w}_{2}, \dots \mathbf{w}_{n_i}\}$ be
the result of performing segmentation on the $i$-th session of a subject,
where $\mathbf{w}_j \in \mathbb{R}^{3 \times W}$ and $n_i$ denotes the
number of segments extracted from session $i$. Given the session segments,
we apply a pre-filtering
step, where only windows with energy above a threshold $E_{\min}$ are kept.
Sessions that end up with less than $2$ segments after this step, are
discarded altogether.
The surviving segments across all sessions are collected in a large pool
and sorted based on their energy in the band $[3, 7]$ Hz (the PD tremor
band). Finally, the top $K_t$ (a scalar hyperparameter, common to all
subjects) segments are drawn 
from the sorted pool and are used to form the bag of the subject, which is
of the form $X = \{\mathbf{x}_1, \mathbf{x}_2, \dots, \mathbf{x}_{K_t}\}$.
A schematic overview of the bag creation procedure for a subject is given
in Figure \ref{bag_creation}.
For subjects that have less than $K_t$ segments, we apply zero-padding to
reach the desired bag length and implement a masking system that ignores
these zero-padded instances during our computations.

\subsection{Model training}
Each extracted bag $X$ is then associated with the available,
expert-provided, tremor label $y$ of the subject,
to form a tuple of the form $(X, y)$. The set of these tuples is the
training set on which we will train our multiple-instance model.

In our previous work \cite{papadopoulos2019mil}, 
the function
$\phi$ operated on
the frequency domain. In particular, we used a fully-connected neural
network to perform feature extraction from the spectral representation of
each instance. In other words, each instance was first transformed into its
frequency representation and then fed into the network.
In this work, we propose an approach in which the function $\phi$ operates
on the time domain and uses a \emph{Convolutional Neural Networks (CNN)} 
to extract features directly from the raw tri-axial acceleration signal.
We do not employ a CNN directly on the spectrogram, because we are
interested in high values in specific spectral coefficients (those
corresponding to $3\hbox{-}7$ Hz) and CNNs are by nature translation invariant.
This means that a filter that has learnt to detect peaks would not be able
to differentiate between a peak at $4$ Hz and a peak at $20$ Hz.
Regarding the rest of the components, the pooling stage $\sigma$ 
is implemented as a simple two-layer fully-connected 
network, while for the final classifier $\rho$ we
use multiple fully-connected layers. Details regarding the specific
architecture choices of each stage are given in Section \ref{experiments}. 

The whole model, comprising the composition of $\rho$, $\sigma$, $\phi$ is
trained end-to-end for $E$ epochs using the cross-entropy loss: 
\begin{equation}
  \begin{split}
  \mathcal{L} = \mathop{-\mathbb{E}}_{X, y \sim \hat{p}_{data}}\big[y
  \log(p\textsubscript{model}(y=1|X)) \\+
  (1-y)(1 - \log(p\textsubscript{model}(y=1|X))\big]
\end{split}
  \label{cross_entropy}
\end{equation}
where $\hat{p}_{data}$ denotes the empirical data distribution defined by
the available training data. Finally, when used for inference, 
the class probability estimate
provided by the model is 
transformed into a class prediction using a threshold $T$.


\begin{table}[!h]
  \centering
  \caption{Demographic characteristics of the subjects in our dataset. For age,
  UPDRS scores, data contribution and relative energy, 
  we report the average value and in parenthesis 
  the standard deviation across the corresponding sub-population. 
  UPDRS20 and UPDRS21 values refer to the sum of both hand scores. 
  Relative energy refers to the ratio of energy in the tremor band (3-7 Hz)
  over the total energy of a segment.}
  \begin{tabular}{lccc}
    \toprule
    & Healthy Controls & PD Patients & Total \\
    \midrule
    Count & 14 & 31 & 45 \\
    Age in years & 55.4 (11.7) & 62.1 (7.3) & 60.0 (9.4)\\
    Years diagnosed & - & 6.3 (3.6) & - \\
    UPDRS 16 & 0.07 (0.25) & 1.09 (0.89) & - \\
    UPDRS 20 & 0 (0) & 1.19 (1.25) & - \\
    UPDRS 21 & 0 (0) & 0.96 (1.51) & - \\
    Sum of UPDRS-III & 2.28 (3.47) & 19.7 (11.5) & - \\
    No. of sessions & 99.6 (151.3) & 131.1 (166.2) & 121.3 (162.4) \\
    Session duration & 30.4 (13.7) & 33.8 (14.6) & 32.9 (14.5) \\
    Processed segments & 441.0 (512.4) & 576.8 (537.2) & 534.6 (533.3) \\
    Relative energy & 0.35 (0.15) & 0.38 (0.21) & 0.37 (0.20) \\
    \bottomrule
  \end{tabular}
\label{demographics}
\end{table}

\section{In-the-wild data collection and pre-processing}\label{dataset}
\subsection{Collection} \label{collection}
We propose to collect in-the-wild IMU data through a mobile application
developed in the context of \cite{lisa_klingelhoefer_2017_1199554}.
The data collection application operates on the background and unobtrusively
initiates recording of the accelerometer sensor
whenever a phone call, either incoming or
outgoing, is placed. The recording lasts at most for the first $75$
seconds of the phone call, 
so as to avoid draining the battery. Each recording, consisting of the
tri-axial accelerometer values, their timestamps and some metadata, is
stored locally and transmitted wirelessly on a server, when the
appropriate conditions, e.g.\ adequate battery levels, Wi-Fi access, etc.,
are met.

\subsection{Pre-processing}\label{preproc}
Due to the wide variety of different recording devices expected during data
collection, a common
pre-processing step is designed and applied to all collected signals to ensure their
homogeneity. First, problematic signals are discarded altogether. The
rejection criteria for a signal are:
\begin{inparaenum}[i)]
  \item its total duration is less than $20$ seconds,
  \item its estimated sampling frequency is below $50$ Hz,
  \item it contains extreme values ($>100\; m/s^2$),
  \item it contains too many missing values.
\end{inparaenum}
Signals that pass the rejection process are subsequently resampled to a
sampling frequency of $100$ Hz, through the use of a polyphase
resampling step that involves linear interpolation, 
a downsampling step of appropriate ratio, a low-pass
filter and an upsampling step, again, of appropriate ratio.
A segment of $5$ seconds is then removed from
the start and the end of the sessions, so as to discard the moments when the
user either picks up or hangs up the phone. In addition, the gravitational
component of the acceleration signal is removed via a high-pass
FIR filter of order $512$ with a cutoff frequency of $1$ Hz.

\subsection{Annotation}\label{annotation}
To acquire tremor annotations for each subject, we resort to
the 
\emph{Unified Parkinson's disease rating scale (UPDRS)} \cite{updrs}.
The UPDRS is the most commonly used scale used by physicians to quantify 
the severity of the various items in the PD symptomatology and keep track
of the disease's longitudinal progression. It includes a self-reported part
(part II), where
subjects themselves provide an estimation of their symptom severity during
daily living,
as well as a motor examination part (part III), 
where the attending physician examines the subject
using a standardized set of motor tests, and provides a clinical evaluation
of their status with respect to each symptom. Each symptom 
(corresponding to a specific UPDRS item) is
given a score, generally
between $0$ and $4$, with $0$ signifying absence of symptom. Regarding tremor, the
relevant UPDRS items are: UPDRS16, the subject's self-report for tremor
at any body part, UPDRS20, the physician's score for hand tremor at rest,
and UPDRS21, the physician's score for action or postural hand tremor. In
reality, UPDRS20 contains separate sub-items concerning the
existence of tremor in each body extremity separately. However, as our
focus is on hand tremor, in the rest of this paper UPDRS20 will refer only
to the hand items.
 
All UPDRS scores that belong to the motor examination part
(UPDRS part III) were obtained by certified neurologists at three
locations: the Department of Neurology of the Technical University of
Dresden, Germany, the Department of Basic and Clinical Neuroscience of the
King's College Hospital, London and the Third Neurological Clinic of the
Papanikolaou Hospital of Thessaloniki. All subjects in the dataset
underwent a thorough neurological
examination, strictly adhering to the protocol of a UPDRS motor
examination, at the location corresponding to their country of residence.

Since our goal is to perform tremor detection,
we use the binarized sum of the individual 
hand scores as the final target label. That is, if the sum of scores across
hands is positive, we consider that the subject belongs to the positive
class (has tremor). Accordingly, if the sum of hand scores is $0$ we consider the
subject to be of the negative class (no tremor). 
This leads to $3$ potential sources
of annotation, mainly the binary versions of UPDRS16, UPDRS20 and UPDRS21.
However, one potential complication is that 
the UPDRS examination provides a clue as to the
tremor severity only at the moment of the examination. This can be problematic
since the examination could have coincided with an off period, and
therefore no tremor would be observed by the doctor. In addition, a subject
may have contributed too few data for the symptom to be observed or they
could have tremor only on one hand and handle their device during phone
calls with the other. Each such eventuality is a source of label noise that could
severely affect the training of our algorithm. Hence, to
overcome these issues we resort to an additional source of annotation that
is produced by a group of signal processing experts 
after visually inspecting the contributed signals of each subject.
A detailed account of how this annotation was produced is given in
\ref{experiments}.

\section{Experimental evaluation}\label{experiments}

In this section, we conduct a series of experiments 
to evaluate our method with respect to the following aspects:
\begin{itemize}
  \item Its efficacy in detecting tremorous episodes in-the-wild.
  \item Its dependence on the bag length.
  \item Its performance relative to popular alternatives.
\end{itemize}

For our experiments, we used a dataset of accelerometer recordings 
that was collected in-the-wild via smartphone-embedded IMU sensors.
More specifically, multiple subjects, both PD patients and healthy
controls, recruited by the medical experts participating in the study, 
downloaded the 
data collection Android application, described in section \ref{collection}, 
and installed it on their personal smartphones.
The typical contribution period of a subject 
ranged from a few weeks to several months, as they were free 
to uninstall the application at any time. Therefore, each subject in the
dataset contributed a different amount of recordings, 
that depended on both the number of phone
calls they realized during the data collection period, as well as the
duration of that period itself.
In addition, we imposed a minimum requirement on the amount of
data contributed from each participant in order for them to be considered.
Specifically, subjects whose bag contained less than $30$ segments in total 
after the session rejection step of section \ref{preproc} and the segment rejection
step of section \ref{bag_creation_section},
were not considered in our
experiments. Ultimately, this led to a final dataset of $45$ subjects.
Additional details regarding the demographics of the subjects and their
data contribution are given in Table \ref{demographics}.

Each subject in the dataset underwent a full clinical evaluation, including
a UPDRS assessment carried out by neurologists, 
at some point during their data contributing period.
In addition, a group of $2$ signal processing experts from our group used the
subjects with the most extreme UPDRS scores to acquire a sense of how
tremor manifests in acceleration signals. Then, the rest of the subjects
were annotated by individually inspecting the raw accelerometer signal of
each recording they had contributed as well as its spectrogram, and taking
into account both the subject's self-report and its clinical evaluation.
During this process, a subject was labeled as tremorous if a tremorous
episode could be detected in at least one of their contributed sessions
upon visual inspection, in compliance with the standard MIL assumption
\cite{dietterich}. The output of this process was used as an extra signal
processing expert annotation.

We evaluated our approach using two experimental methodologies: 
a \emph{Leave One Subject Out (LOSO)} scheme and a \emph{Repeated k-Fold
(RkF)} scheme.
In the LOSO scheme, we use the data from all the subjects except one
to train a model, evaluate the trained model on the left-out subject and
repeat the process so that all subjects are used for evaluation once.
In the RkF scheme, we split the subject group into $k$ parts, use the data
from the subjects belonging to the $k-1$ parts to train a model 
and evaluate it on the subjects belonging to the other part. 
The process is repeated so that all parts are used once for
evaluation and multiple repetitions of this procedure are conducted (hence,
the ``repeated'' in its name) for different permutations of the subject group.
In each repetition of both experimental schemes, we conducted multiple 
trials to account for the randomness inherent
in the examined algorithms. We performed training using the signal processing
experts annotations, as we consider them to be the most robust. Evaluation
was performed using each of the available annotations, mainly UPDRS16, UPDRS20,
UPDRS21 and signal processing expert (denoted SP-expert for short)
annotations.

For computing the energy of a segment in the band of $3-7$Hz (a step 
required during the bag creation phase) we used its frequency
representation (computed according to Welch's method for spectral density
estimation, as described below)
and summed the components that corresponded to that frequency band. We also
used $E_{\min} = 0.15$ for discarding segments with low activity.
For our deep multiple-instance approach,
we used a window length $W$ of $500$ samples, i.e. $5$ seconds given the
common sampling frequency of $100$ Hz. 
For the instance embedding function, $\phi$, we examined 2 different
approaches, the one proposed in \cite{papadopoulos2019mil} 
that operates on the frequency domain and the one proposed in this paper that
operates on the time domain. The frequency domain approach, 
transforms each acceleration axis individually to the 
frequency domain using Welch's method with a window size of $3$ seconds and
$75\%$ overlap. Subsequently, the spectra across the 3 axes are summed and the
coefficients that correspond to the frequency range $[0, 25]$ Hz are kept,
thus leading to a vector of $76$ elements. This frequency representation
is then fed to a fully-connected network of multiple layers to acquire
the instance embedding.
On the contrary, the time domain approach operates directly on
raw signal values and employs a CNN to extract features for
each instance. 
In the following, we will refer to the former approach as
\emph{Deep-MIL-FC} and
to the latter as \emph{Deep-MIL-CNN}.

For both approaches, we resorted to standard network architectures that
are used throughout the literature for classification problems.
Specific details about each architecture are given in Table
\ref{extractor}. 
Apart from the embedding function $\phi$, we kept the rest of the
model architecture identical.
We used an embedding dimension, $M$, equal to $64$.
The attention pooling stage was modeled by two
fully-connected layers that implemented Equation \ref{attention}, with the
attention dimension, $L$, set to $16$. The values for the 
hyperparameters $M$, $L$ were
selected by experimentation on a small subset of the dataset.
The final classifier stage, $\rho$,
was also implemented as a fully-connected network with multiple layers
(described in Table \ref{clf}).
The whole model was trained end-to-end for
$E$ epochs using the Adam \cite{kingma2014adam} optimizer with learning
rate $\epsilon=0.001$ and the suggested default values 
for the parameters $\beta_1, \beta_2$. Learning rate decay was also applied
during the last half of the training that exponentially reduced the 
learning rate at the beginning of each epoch by a factor of $0.9$.
The decision threshold, $T$, was set to $0.5$.
In total, the Deep-MIL-CNN model had $46627$ 
trainable parameters and the Deep-MIL-FC model $65603$.

\begin{table*}[!h]
  \centering
  \caption{Evaluation results for the LOSO experiment with bag
    length $K_t=1500$. For each method we report the 
    average and the standard deviation of its performance 
    metrics across 10 independent LOSO trials. 
    Notice that the implementation of MI-SVM that 
    we use is fully deterministic, 
    so there is no variance
    in its performance.
    Deep-MIL-FC was trained for
  $E=1000$ epochs and Deep-MIL-CNN for $E=50$ epochs. UPDRS annotations refer 
to binary versions of the physician-provided UPDRS scores.}
  \begin{tabular}{clcccc}
    \toprule
    \textbf{Evaluation on} & \textbf{Model} & \textbf{Precision} & \textbf{Sensitivity} &
    \textbf{Specificity} & \textbf{F1-score} \\
    \midrule
    \multirow{10}{*}{\parbox{4cm}{\centering UPDRS16 \\ (24 positive - 21
    negative)}}
    & Simple-MIL & 0.862 $\pm$ 0.047 & 0.362 $\pm$ 0.027 & 0.933 $\pm$
    0.023 & 0.522 $\pm$ 0.030 \\
    & MI-Net-Simple \cite{minet} & 0.808 $\pm$ 0.052 & 0.442 $\pm$ 0.028 &
    0.876 $\pm$ 0.044 & 0.586 $\pm$ 0.020 \\
    & MI-Net-DS \cite{minet} & 0.759 $\pm$ 0.036 & 0.454 $\pm$ 0.012 & 0.833 $\pm$
    0.032 & 0.588 $\pm$ 0.012 \\
    & MI-Net-Res \cite{minet} & 0.711 $\pm$ 0.034 & 0.458 $\pm$ 0.019 &
    0.786 $\pm$ 0.032 & 0.579 $\pm$ 0.017 \\
    & BSN \cite{bagrep} & 0.801 $\pm$ 0.054 & 0.425 $\pm$ 0.045 & 0.876 $\pm$
    0.049 & 0.570 $\pm$ 0.041 \\
    & Deep-MIL-FC \cite{papadopoulos2019mil}& 0.799 $\pm$ 0.004 & \textbf{0.496} $\pm$
    0.013 & 0.857 $\pm$ 0.001 & \textbf{0.628} $\pm$ 0.010 \\
    & Deep-MIL-CNN (this paper)& 0.780 $\pm$ 0.028 & 0.475 $\pm$ 0.038 
    & 0.848 $\pm$ 0.019 & 0.608 $\pm$ 0.033 \\
    &MI-SVM \cite{andrews2003support}& \textbf{1.000} $\pm$ 0.000 & 0.250 $\pm$ 0.000 &
    \textbf{1.000} $\pm$ 0.000 & 0.400  $\pm$ 0.000\\
    &BoF + SVM \cite{wei2016scalable}& 0.782 $\pm$ 0.098 & 0.208 $\pm$ 0.019 & 0.929 $\pm$ 0.038
    & 0.340 $\pm$ 0.024 \\
    &FV + SVM \cite{wei2016scalable}& 0.597 $\pm$ 0.036 & 0.438 $\pm$ 0.057 & 0.662 $\pm$ 0.050 &
    0.523 $\pm$ 0.036 \\
    \midrule
    \multirow{10}{*}{\parbox{4cm}{\centering UPDRS20 \\ (17 positive - 28
    negative)}}
    & Simple-MIL & \textbf{0.734} $\pm$ 0.054 & 0.435 $\pm$ 0.029 & 0.904 $\pm$
    0.023 & 0.587 $\pm$ 0.029 \\
    & MI-Net-Simple \cite{minet} & 0.620 $\pm$ 0.060 & 0.476 $\pm$ 0.018 &
    0.818 $\pm$ 0.046 & 0.601 $\pm$ 0.016 \\
    & MI-Net-DS \cite{minet} & 0.597 $\pm$ 0.014 & 0.506 $\pm$ 0.029 &
    0.793 $\pm$ 0.014 & 0.617 $\pm$ 0.019 \\
    & MI-Net-Res \cite{minet} & 0.568 $\pm$ 0.024 & 0.518 $\pm$ 0.024 &
    0.761 $\pm$ 0.023 & 0.616 $\pm$ 0.018 \\
    & BSN \cite{bagrep} & 0.551 $\pm$ 0.037 & 0.412 $\pm$ 0.037 & 0.793 $\pm$
    0.042 & 0.540 $\pm$ 0.027 \\
    &Deep-MIL-FC \cite{papadopoulos2019mil}& 0.731 $\pm$ 0.006 & \textbf{0.641} $\pm$ 0.018 & 0.857 $\pm$ 0.001 &
    \textbf{0.733} $\pm$ 0.012 \\
    &Deep-MIL-CNN (this paper)& 0.714 $\pm$ 0.048 & 0.612 $\pm$ 0.039 & 0.850 $\pm$ 0.031 &
    0.711 $\pm$ 0.032 \\
    &MI-SVM \cite{andrews2003support}& 0.667 $\pm$ 0.000 & 0.235 $\pm$ 0.000 & 0.929 $\pm$ 0.000 & 0.375 $\pm$ 0.000 \\
    &BoF + SVM \cite{wei2016scalable}& 0.725 $\pm$ 0.077 & 0.276 $\pm$ 0.046  & \textbf{0.936} $\pm$ 0.021
    & 0.425 $\pm$ 0.056 \\
    &FV + SVM \cite{wei2016scalable}& 0.473 $\pm$ 0.033 & 0.488 $\pm$ 0.053 & 0.668 $\pm$ 0.051 &
    0.560 $\pm$ 0.029 \\
    \midrule
    \multirow{10}{*}{\parbox{4cm}{\centering UPDRS21 \\ (13 positive - 32
    negative)}}
    & Simple-MIL & 0.575 $\pm$ 0.034 & 0.446 $\pm$ 0.031 & 0.866 $\pm$
    0.014 & 0.588 $\pm$ 0.029 \\
    & MI-Net-Simple \cite{minet} & 0.466 $\pm$ 0.045 & 0.469 $\pm$ 0.023 &
    0.778 $\pm$ 0.041 & 0.585 $\pm$ 0.018 \\
    & MI-Net-DS \cite{minet} & 0.458 $\pm$ 0.015 & 0.508 $\pm$ 0.038 &
    0.756 $\pm$ 0.012 & 0.606 $\pm$ 0.019 \\
    & MI-Net-Res \cite{minet} & 0.439 $\pm$ 0.022 & 0.523 $\pm$ 0.031 &
    0.728 $\pm$ 0.020 & 0.608 $\pm$ 0.022 \\
    & BSN \cite{bagrep} & 0.392 $\pm$ 0.026 & 0.385 $\pm$ 0.049 & 0.756 $\pm$
    0.036 & 0.507 $\pm$ 0.038 \\
    &Deep-MIL-FC \cite{papadopoulos2019mil}& \textbf{0.597} $\pm$ 0.009 & \textbf{0.685} $\pm$ 0.023 & 0.812 $\pm$ 0.001 &
    \textbf{0.743} $\pm$ 0.014 \\
    &Deep-MIL-CNN (this paper)& 0.576 $\pm$ 0.044 & 0.646 $\pm$ 0.051 & 0.806 $\pm$ 0.027 &
    0.716 $\pm$ 0.037 \\
    &MI-SVM \cite{andrews2003support}& 0.333 $\pm$ 0.000 & 0.154 $\pm$ 0.000 & 0.875 $\pm$ 0.000 & 0.262 $\pm$ 0.000 \\
    &BoF + SVM \cite{wei2016scalable}& 0.567 $\pm$ 0.078 & 0.285 $\pm$ 0.060  & \textbf{0.912} $\pm$ 0.019
    & 0.430 $\pm$ 0.073 \\
    &FV + SVM \cite{wei2016scalable}& 0.411 $\pm$ 0.031 & 0.554 $\pm$ 0.058 & 0.675 $\pm$ 0.049&
    0.605 $\pm$ 0.030 \\
    \midrule
    \multirow{10}{*}{\parbox{4cm}{\centering SP-expert annotations\\ (16
    positive - 29 negative)}}
    & Simple-MIL & 0.753 $\pm$ 0.044 & 0.475 $\pm$ 0.031 & 0.914 $\pm$
    0.017 & 0.625 $\pm$ 0.029 \\
    & MI-Net-Simple \cite{minet} & 0.808 $\pm$ 0.052 & 0.662 $\pm$ 0.041 &
    0.910 $\pm$ 0.032 & 0.765 $\pm$ 0.023 \\
    & MI-Net-DS \cite{minet} & 0.799 $\pm$ 0.019 & 0.719 $\pm$ 0.042 &
    0.900 $\pm$ 0.010 & 0.798 $\pm$ 0.026 \\
    & MI-Net-Res \cite{minet} & 0.756 $\pm$ 0.029 & 0.731 $\pm$ 0.029 &
    0.869 $\pm$ 0.021 & 0.794 $\pm$ 0.018 \\
    & BSN \cite{bagrep} & 0.787 $\pm$ 0.062 & 0.625 $\pm$ 0.056 & 0.903 $\pm$
    0.037 & 0.737 $\pm$ 0.036 \\
    &Deep-MIL-FC \cite{papadopoulos2019mil}& 0.933 $\pm$ 0.001 & 0.869 $\pm$ 0.019 & 0.966 $\pm$ 0.001 &
    0.914 $\pm$ 0.011 \\
    &Deep-MIL-CNN (this paper)& \textbf{0.987} $\pm$ 0.027 & \textbf{0.900} $\pm$ 0.057
    & \textbf{0.993} $\pm$ 0.014 &
    \textbf{0.943} $\pm$ 0.034 \\
    &MI-SVM \cite{andrews2003support}& 0.833 $\pm$ 0.000 & 0.312 $\pm$ 0.000 & 0.966 $\pm$ 0.000 & 0.472 $\pm$ 0.000 \\
    &BoF + SVM \cite{wei2016scalable}& 0.883 $\pm$ 0.084 & 0.356 $\pm$ 0.049 & 0.972 $\pm$ 0.021
    & 0.519 $\pm$ 0.054 \\
    &FV + SVM \cite{wei2016scalable}& 0.580 $\pm$ 0.071 & 0.631 $\pm$ 0.059 & 0.741 $\pm$ 0.064 &
    0.679 $\pm$ 0.043 \\
    \bottomrule
  \end{tabular}
\label{results_loso}
\end{table*}

\begin{table*}[!h]
  \centering
  \caption{Evaluation results for the Repeated $k$-Fold experiment
    with $k=5$ and bag length $K_t=1500$. 
    For each method we report the 
    average and the standard deviation of its performance 
    metrics across 10 repetitions (and 5 random trials per repetition) 
    of the 5-fold experiment. 
}
  \begin{tabular}{clcccc}
    \toprule
    \textbf{Evaluation on} & \textbf{Model} & \textbf{Precision} & \textbf{Sensitivity} &
    \textbf{Specificity} & \textbf{F1-score} \\
    \midrule
    \multirow{9}{*}{\parbox{4cm}{\centering UPDRS16 \\ (24 positive - 21
    negative)}}
    & Simple-MIL & \textbf{0.846} $\pm$ 0.017 & 0.356 $\pm$ 0.006 & \textbf{0.923} $\pm$
    0.010 & 0.514 $\pm$ 0.006 \\
    & MI-Net-Simple \cite{minet} & 0.827 $\pm$ 0.016 & 0.470 $\pm$ 0.010 &
    0.888 $\pm$ 0.010 & \textbf{0.615} $\pm$ 0.011 \\
    & MI-Net-DS \cite{minet} & 0.754 $\pm$ 0.006 & \textbf{0.479} $\pm$ 0.004 &
    0.821 $\pm$ 0.006 & 0.605 $\pm$ 0.003 \\
    & MI-Net-Res \cite{minet} & 0.751 $\pm$ 0.012 & 0.478 $\pm$ 0.011 &
    0.819 $\pm$ 0.009 & 0.604 $\pm$ 0.010 \\
    & BSN \cite{bagrep} & 0.802 $\pm$ 0.021 & 0.426 $\pm$ 0.013 & 0.880 $\pm$
    0.014 & 0.574 $\pm$ 0.013 \\
    & Deep-MIL-FC \cite{papadopoulos2019mil}& 0.803 $\pm$ 0.007 &
    0.474 $\pm$ 0.003 & 0.867 $\pm$ 0.006 & 0.613 $\pm$ 0.003 \\
    & Deep-MIL-CNN (this paper)& 0.778 $\pm$ 0.012 & 0.449 $\pm$ 0.009 &
    0.853 $\pm$ 0.010 & 0.588 $\pm$ 0.008 \\
    &BoF + SVM \cite{wei2016scalable}& 0.754 $\pm$ 0.027 & 0.247 $\pm$
    0.007 & 0.908 $\pm$ 0.013 & 0.388 $\pm$ 0.008 \\
    &FV + SVM \cite{wei2016scalable}& 0.627 $\pm$ 0.011 & 0.460 $\pm$ 0.011
    & 0.688 $\pm$ 0.014 &
    0.551 $\pm$ 0.008 \\
    \midrule
    \multirow{9}{*}{\parbox{4cm}{\centering UPDRS20 \\ (17 positive - 28
    negative)}}
    & Simple-MIL & \textbf{0.759} $\pm$ 0.015 & 0.457 $\pm$ 0.008 & 0.856 $\pm$
    0.008 & 0.595 $\pm$ 0.009 \\
    & MI-Net-Simple \cite{minet} & 0.617 $\pm$ 0.010 & 0.495 $\pm$ 0.007 &
    0.814 $\pm$ 0.005 & 0.616 $\pm$ 0.007 \\
    & MI-Net-DS \cite{minet} & 0.554 $\pm$ 0.004 & 0.498 $\pm$ 0.003 &
    0.757 $\pm$ 0.005 & 0.601 $\pm$ 0.002 \\
    & MI-Net-Res \cite{minet} & 0.559 $\pm$ 0.004 & 0.502 $\pm$ 0.006 &
    0.759 $\pm$ 0.005 & 0.605 $\pm$ 0.004 \\
    & BSN \cite{bagrep} & 0.564 $\pm$ 0.008 & 0.422 $\pm$ 0.007 & 0.801 $\pm$
    0.009 & 0.553 $\pm$ 0.004 \\
    &Deep-MIL-FC \cite{papadopoulos2019mil}& 0.719 $\pm$ 0.006 &
    \textbf{0.600} $\pm$ 0.004 & 0.858 $\pm$ 0.004 & \textbf{0.706} $\pm$ 0.003 \\
    &Deep-MIL-CNN (this paper)& 0.723 $\pm$ 0.012 & 0.589 $\pm$ 0.004 &
    0.863 $\pm$ 0.009 &
    0.700 $\pm$ 0.002 \\
    &BoF + SVM \cite{wei2016scalable}& 0.682 $\pm$ 0.019 & 0.315 $\pm$
    0.013  & \textbf{0.911} $\pm$ 0.007
    & 0.468 $\pm$ 0.015 \\
    &FV + SVM \cite{wei2016scalable}& 0.512 $\pm$ 0.013 & 0.529 $\pm$ 0.004
    & 0.693 $\pm$ 0.016 &
    0.600 $\pm$ 0.006 \\
    \midrule
    \multirow{9}{*}{\parbox{4cm}{\centering UPDRS21 \\ (13 positive - 32
    negative)}}
    & Simple-MIL & 0.542 $\pm$ 0.015 & 0.456 $\pm$ 0.009 & 0.856 $\pm$
    0.008 & 0.595 $\pm$ 0.009 \\
    & MI-Net-Simple \cite{minet} & 0.471 $\pm$ 0.010 & 0.494 $\pm$ 0.009 &
    0.774 $\pm$ 0.005 & 0.603 $\pm$ 0.008 \\
    & MI-Net-DS \cite{minet} & 0.423 $\pm$ 0.003 & 0.497 $\pm$ 0.004 &
    0.725 $\pm$ 0.004 & 0.590 $\pm$ 0.002 \\
    & MI-Net-Res \cite{minet} & 0.428 $\pm$ 0.004 & 0.503 $\pm$ 0.008 &
    0.727 $\pm$ 0.004 & 0.595 $\pm$ 0.005 \\
    & BSN \cite{bagrep} & 0.407 $\pm$ 0.005 & 0.398 $\pm$ 0.009 & 0.764 $\pm$
    0.008 & 0.524 $\pm$ 0.006 \\
    & Deep-MIL-FC \cite{papadopoulos2019mil}& 0.578 $\pm$ 0.005 &
    \textbf{0.631} $\pm$ 0.005 & 0.813 $\pm$ 0.004 & \textbf{0.710} $\pm$ 0.003 \\
    & Deep-MIL-CNN (this paper)& \textbf{0.579} $\pm$ 0.009 & 0.617 $\pm$ 0.006 &
    0.818 $\pm$ 0.008 & 0.703 $\pm$ 0.003 \\
    &BoF + SVM \cite{wei2016scalable}& 0.555 $\pm$ 0.019 & 0.335 $\pm$
    0.017  & \textbf{0.891} $\pm$ 0.006
    & 0.487 $\pm$ 0.018 \\
    &FV + SVM \cite{wei2016scalable}& 0.425 $\pm$ 0.011 & 0.575 $\pm$ 0.011
    & 0.684 $\pm$ 0.013&
    0.625 $\pm$ 0.009 \\
    \midrule
    \multirow{9}{*}{\parbox{4cm}{\centering SP-expert annotations\\ (16
    positive - 29 negative)}}
    & Simple-MIL & 0.766 $\pm$ 0.017 & 0.528 $\pm$ 0.005 & 0.920 $\pm$
    0.008 & 0.671 $\pm$ 0.004 \\
    & MI-Net-Simple \cite{minet} & 0.820 $\pm$ 0.006 & 0.699 $\pm$ 0.005 &
    0.915 $\pm$ 0.003 & 0.792 $\pm$ 0.004 \\
    & MI-Net-DS \cite{minet} & 0.744 $\pm$ 0.003 & 0.710 $\pm$ 0.008 &
    0.866 $\pm$ 0.002 & 0.780 $\pm$ 0.005 \\
    & MI-Net-Res \cite{minet} & 0.754 $\pm$ 0.007 & 0.720 $\pm$ 0.014 &
    0.870 $\pm$ 0.004 & 0.788 $\pm$ 0.009 \\
    & BSN \cite{bagrep} & 0.787 $\pm$ 0.020 & 0.626 $\pm$ 0.005 & 0.906 $\pm$
    0.011 & 0.741 $\pm$ 0.005 \\
    &Deep-MIL-FC \cite{papadopoulos2019mil}& 0.913 $\pm$ 0.003 & 0.809
    $\pm$ 0.006 & 0.957 $\pm$ 0.002 &
    0.877 $\pm$ 0.003 \\
    &Deep-MIL-CNN (this paper)& \textbf{0.955} $\pm$ 0.012 & \textbf{0.828}
    $\pm$ 0.024
    & \textbf{0.979} $\pm$ 0.006 &
    \textbf{0.897} $\pm$ 0.015 \\
    &BoF + SVM \cite{wei2016scalable}& 0.812 $\pm$ 0.021 & 0.399 $\pm$
    0.016 & 0.949 $\pm$ 0.006
    & 0.561 $\pm$ 0.016 \\
    &FV + SVM \cite{wei2016scalable}& 0.722 $\pm$ 0.017 & 0.684 $\pm$ 0.008
    & 0.770 $\pm$ 0.016 &
    0.724 $\pm$ 0.009 \\
    \bottomrule
  \end{tabular}
\label{results_5fold}
\end{table*}

\begin{table}[!h]
  \centering
  \caption{The architectures used for the embedding function $\phi$. $k$
  denotes the kernel size, $f$ the number of filters in the convolutional
  layers and $M$ the final embedding dimension.}
  \begin{tabular}{ccc}
    \toprule
     & \textbf{Fully-connected} & \textbf{CNN} \\
    \midrule
    Input $\mathbf{x}_k$ & $1\times76$ spectrogram & $3\times500$ raw acceleration \\
    \midrule
     \multirow{3}{*}{Layer 1}
     & Dense $76 \to 256$ & Conv1D $k=8, f=32$ \\ 
     & Leaky-ReLU ($\alpha = 0.2$) & Leaky-ReLU ($\alpha = 0.2$) \\
     & Dropout $p=0.5$ & MaxPool $k=2$ \\
    \midrule
     \multirow{3}{*}{Layer 2}
     & Dense $256 \to 128$ & Conv1D $k=8, f=32$ \\ 
     & Leaky-ReLU ($\alpha = 0.2$) & Leaky-ReLU ($\alpha = 0.2$) \\
     & Dropout $p=0.5$ & MaxPool $k=2$ \\
    \midrule
     \multirow{3}{*}{Layer 3}
     & \multirow{3}{*}{Dense $128 \to M$} & Conv1D $k=16, f=16$ \\
     &  & Leaky-ReLU ($\alpha = 0.2$) \\
     & & MaxPool $k=2$ \\ 
    \midrule
     \multirow{3}{*}{Layer 4}
     &           & Conv1D $k=16, f=16$ \\
     &           & Leaky-ReLU ($\alpha = 0.2$) \\
     &           & MaxPool $k=2$\\ 
    \midrule
    \multirow{2}{*}{Layer 5}
    &  & Flatten \\
    &  & Dense $320 \to M$ \\
    \midrule
    Output & $\mathbf{h}_k \in \mathcal{R}^M$ & $\mathbf{h}_k \in
    \mathcal{R}^M$ \\
    \bottomrule
  \end{tabular}
\label{extractor}
\end{table}

\begin{table}[!h]
  \centering
  \caption{Architecture of the final classifier $\rho$. $M$ denotes the
  output dimension of the instance embedding function, $\phi$.}
  \begin{tabular}{cc}
    \toprule
    & \textbf{Final classifier stage} \\
    \midrule
    Input & $\mathbf{z} \in \mathcal{R}^M$ (see Equation \ref{pooling}) \\
    \midrule
     \multirow{3}{*}{Layer 1}
     & Dense $M \to 32$ \\ 
     & Leaky-ReLU ($\alpha = 0.2$)\\ 
     & Dropout $p=0.2$ \\
    \midrule
     \multirow{3}{*}{Layer 2}
     & Dense $32 \to 16$ \\
     & Leaky-ReLU ($\alpha = 0.2$)\\
     & Dropout $p=0.2$\\
    \midrule
     \multirow{2}{*}{Layer 3}
     & Dense $16 \to 2$ \\
     & 2-way softmax\\
    \midrule
     Output & $p(y|X)$ \\
    \bottomrule
  \end{tabular}
\label{clf}
\end{table}

We compared our method against $7$ 
alternatives for solving MIL problems. 
More specifically, we employed a \emph{Bag of Features
(BoF)} \cite{csurka2004visual} scheme for encoding a subject's bag and then
used an SVM with the chi-square kernel to train a classifier on the 
resulting bag encodings. 
We also examined the use of the more robust \emph{Fisher Vector (FV)}
\cite{sanchez2013image} encoding
scheme coupled with a linear SVM, similar in spirit to
\cite{wei2016scalable}. We also compared against the \emph{Multiple Instance
SVM (MI-SVM)} algorithm \cite{andrews2003support}, an SVM variant whose formulation
is modified so that it can solve multiple-instance problems. In
addition, we evaluated against the models proposed by the authors of
\cite{minet}, who suggest a similar
architecture to the methodology adopted in this paper. That is, a
feature-extraction network that operates on each bag instance
independently, followed by a pre-defined pooling operator, such as mean, max or
log-sum-exp,
and, ultimately, a linear transformation to the bag label probability.  
They propose 3 different models for producing the bag embedding: 
a fully-connected network followed by the pooling operator, 
a fully-connected network with deep supervision, where
for each hidden layer a different bag pooling is produced and used for
classification,
and a fully-connected network with residual connections, where in each
hidden layer the pooling operator is applied on the 
sum of the current hidden representation
with the hidden representation of the previous layer.
Lastly, we compared with the approach of \cite{bagrep}, that builds upon the 
MI-Net model to introduce
a 2 stage approach: it first
learns a bag similarity metric via a trained MI-Net model
and it then uses it to derive a bag
representation vector, based on the similarities of the given bag 
with a set of reference bags.

We denote the above models as
\emph{MI-Net-Simple}, \emph{MI-Net-DS}, \emph{MI-Net-Res} and \emph{BSN}
respectively.
For each of these methods, we used max pooling as the pooling operator, 
because it was reported in both works
to give consistent performance across different problems.
Finally, as a very simple baseline, we
also compared against a ``naive'' MIL algorithm (denoted as Simple-MIL in
Table \ref{results_loso}), in which the bag label (i.e
the subject label) was propagated to all the bag instances and a standard
supervised model that classifies segments was trained. The decision for the
left-out
subject was then computed as the average of the model's predictions for all
the instances in their bag. 

For efficiency purposes, all 
the algorithms used for comparison 
(including the
simple-MIL approach, which used the exact same architecture as 
the Deep-MIL-FC algorithm)
operated on the frequency domain, that is, they accepted as
input bags of spectrograms, computed in the same way as in the 
frequency-based MIL approach.
For the BoF encoding
we used a codebook of size $128$, while for the FV encoding we used a 
GMM with $64$ modes. In both cases, the base $C$ hyperparameter of 
the SVM was set to 1 and then balanced according to the class prior.
For the MI-SVM approach, we used a linear kernel with $C$ equal to $100$, after 
experimenting with as small subset of the data. Finally, for comparing
  against
  the MI-Net variants, we used an underlying network architecture as close as
possible to the Deep-MIL-FC algorithm.

Table \ref{results_loso} presents the results of the LOSO experiment for
our approach, as well as
the alternative algorithms of the previous paragraph for
each available evaluation scheme. Table \ref{results_5fold}
presents the respective results for the RkF experiment.
We can see that the attention-based models outperform the
alternatives under almost all evaluation schemes. Specifically, when
evaluation is performed on the signal processing expert labels, 
which can be considered as the most
correct, our approach outperforms the alternatives by a significant margin. 
Overall, the Deep-MIL-CNN model achieves the best performance
suggesting that it is feasible to obtain good performance by training CNN
feature extractors from scratch under the weakly-labeled scenario of
multiple-instance learning. The observed gap in performance 
between the signal processing expert annotations and the medical expert ones, 
may appear large but the observed discrepancy 
can be attributed to the occurrence of label noise, 
the causes of which are
discussed in the last paragraph of Section \ref{annotation}.

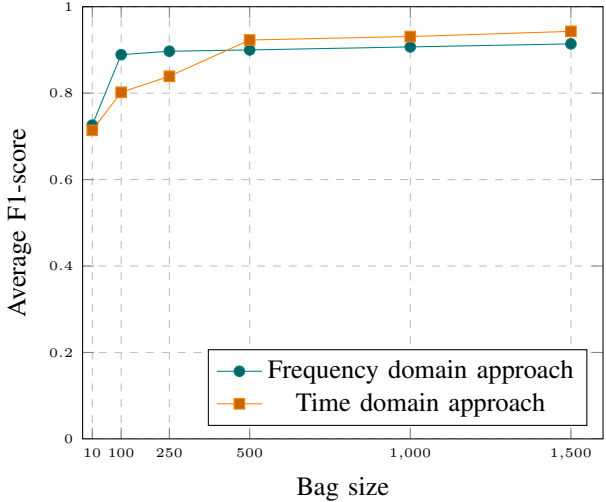
\begin{figure}
  \centering
\begin{tikzpicture}
\begin{axis}[
    xlabel={Bag size},
    ylabel={Average F1-score},
    xmin=-20, xmax=1600,
    ymin=0, ymax=1,
    xtick={10, 100, 250, 500, 1000, 1500},
    legend pos=south east,
    ymajorgrids=true,
    xmajorgrids=true,
    grid style=dashed,
    cycle list name=exotic,
]

\addplot+[
    ]
    coordinates {
      (10,0.726)(100,0.889)(250,0.897)(500, 0.9)(1000, 0.907)(1500, 0.914)
    };
    \addlegendentry{Frequency domain approach}

\addplot+[
    ]
    coordinates {
      (10, 0.714)(100, 0.802)(250,0.839)(500, 0.923)(1000, 0.931)(1500, 0.943)
    };
    \addlegendentry{Time domain approach}

\end{axis}
\end{tikzpicture}
\caption{Model performance (measured by the average F1-score across $10$ LOSO
trials) as a function of the bag size}
\label{bag_size}
\end{figure}

We also examine how different bag size values affect the
classification performance. To this end we repeat the same experiment as
above while varying the bag size. The change in performance as a 
function of the bag size $K_t$ can be seen
in Figure \ref{bag_size}. In general, we can see that reasonable performance can be
achieved with as little as $100$ instances per bag. Moreover, it is
interesting to notice that at the low bag size regime, the frequency domain
approach achieves the best performance. This can be explained by the
fact that CNNs require a large amount of data to be trained efficiently.
Thus, as the bag size increases, more data are presented to the CNN and 
so the time domain approach catches up and
finally outperforms the frequency-based approach.

\begin{figure}[!ht]
  \centering
  \includegraphics[width=1\linewidth]{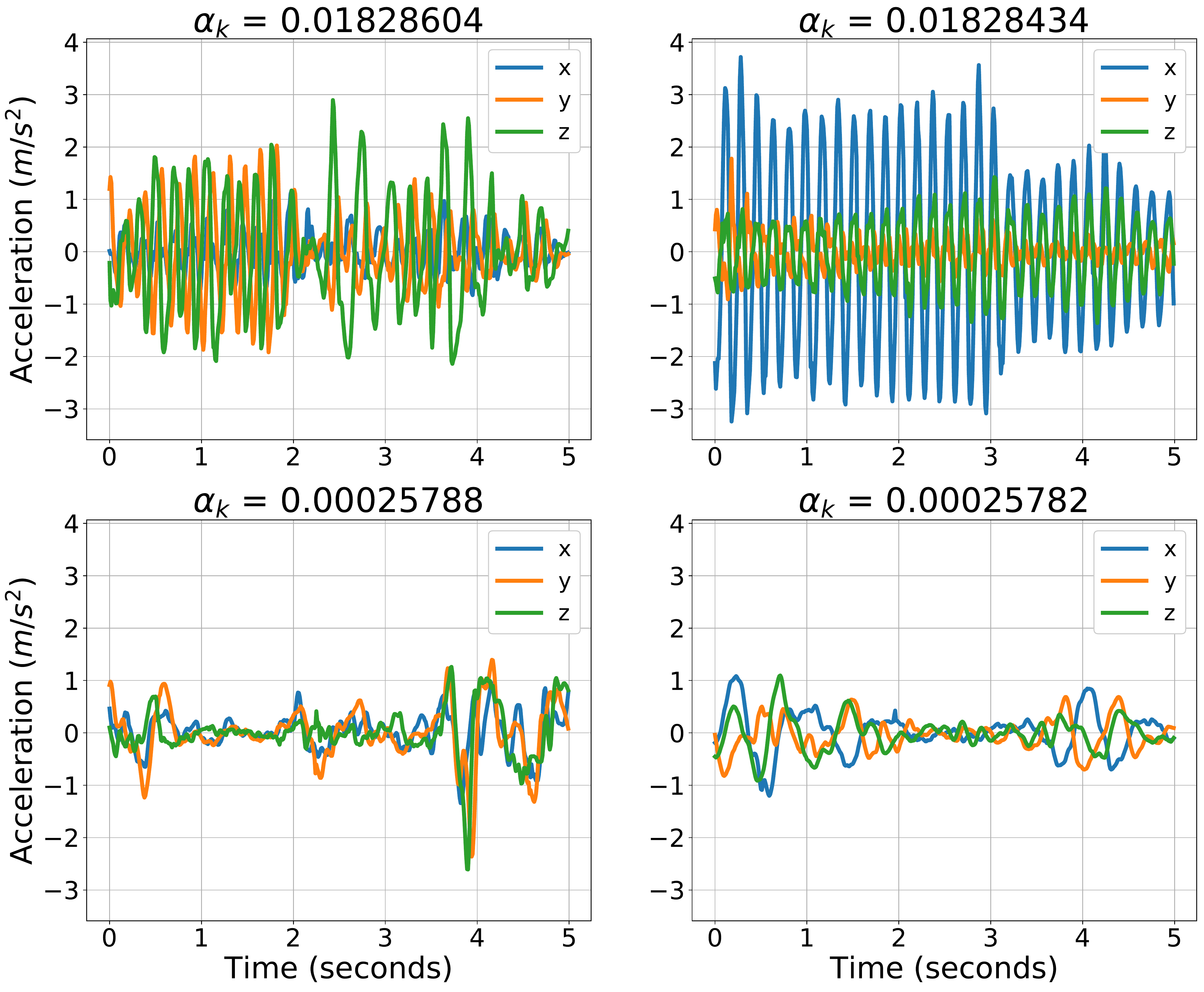}
  \caption{Visualization of the top-2 and bottom-2 instances and their
    corresponding $a_k$ coefficients, as identified by the CNN-based
    model for a tremorous subject. TOP ROW: The 2 instances
      with the largest $a_k$ for the given subject. BOTTOM ROW: The 2
      instances with the lowest $a_k$ for the given subject.
  }
  \label{top_bottom}
\end{figure}

Finally, we wish to evaluate the ability of the attention-based model to discover
key instances within a bag. To that end, 
we visualize the $2$ instances with the highest
$\alpha_k$ and the $2$ instances with the lowest $\alpha_k$, as
identified by the Deep-MIL-CNN model for a tremorous subject.
As we can see in Figure \ref{top_bottom}, the model correctly assigns large
weights to instances that contain tremorous episodes (sinusoidal components
of $\approx 7Hz$) and low weights to instances without such patterns.
This corroborates our hope that throughout its training, the model learns to
discover tremor-related patterns in large sets of heterogeneous (with
respect to the class label) instances, without 
being explicitly presented with
such patterns, as in traditional supervised learning.

\section[]{Discussion}\label{discussion}

In general, we can say that the
proposed method exhibits quite high classification performance and clearly
outperforms the alternative methods.
One explanation for this, is that the attention-based pooling scheme
can accurately identify the
positive instances within a bag. In problems where the data
collection takes place in-the-wild and the targeted symptom is
of intermittent nature,
the number of positive
instances within a bag can be very small. Thus, their contribution
to the bag representation 
may be lost when performing frequency-based pooling (as
in BoW or FV encoding) or mean pooling.
On the contrary, the positive instances in
the case of a well-trained attention-based pooling scheme will receive
high weight, thus contributing significantly to the final bag
representation. Combining the attention mechanism to a robust CNN feature
extractor can, therefore, lead to the very good detection results we report.
As a final note on this, we conjecture that the max 
pooling operator should perform equally well, since, in theory, 
it is insensitive to the small number of positive instances.
However, we can
see in the experimental results that this is not the case, suggesting that the
attention-based pooling is more suitable for the training process.

However, in a real world deployment, the method would face a series of
challenges, owing to the particular nature of tremor and the consequently
problematic labeling mechanism.
First of all, for evaluation purposes in such a scenario, 
any predictions made by the model would be compared against the
opinions of the medical experts. 
If these opinions are expressed in terms of a single UPDRS evaluation,
the performance of the algorithm will be underestimated, since, as we saw
in Section \ref{annotation}, a single UPDRS examination is prone to
producing erroneous labels, due to the tremor's intermittence.
Hence, some of the model's predictions, while in fact correct, 
would be perceived as wrong. A
possible countermeasure to this, would be for the medical experts 
to re-evaluate these ``false positive'' predictions.
In such a scenario, the number of subjects who will ``waist'' such a 
trip to the doctor, is expected to be very low, since the 
specificity of the model, as estimated by our experiments, is very high
($>99\%$). High specificity is a very desirable 
property for an algorithm that operates as a warning system within a mobile
phone environment and therefore aims for
unobtrusiveness, because, in such use
cases, a high
false-positive rate is considered to be more detrimental 
than a high false-negative rate.

A different kind of problem appears when a subject has tremor in only
one hand, but uses the unaffected hand to answer their phone calls.
In such cases, we cannot hope to make a successful prediction, since all
the data contributed by the subject will not contain tremorous episodes.
Yet, according to the medical experts, the subject will belong to the
positive class. Thus, this would unavoidably result in a false negative
prediction.
This is also true for cases where the subject makes a call with a bluetooth set, 
es in the current version of the data collection app
we do not record whether bluetooth was used during the call and therefore
cannot filter it out.

Another point of concern, stems from the dataset itself. 
As we can see in Table \ref{demographics}, the
ratio of Healthy controls over PD patients in our dataset 
(14 Healthy and 31 PD), is very different from
the expected ratio in the general population (about $1\%$ of the
population above 60 years old). This is a result of the subject recruitment
process, conducted by the medical experts participating in the study.
In future work, we plan to
re-evaluate our method on a larger dataset with many more subjects, 
so as to mitigate this concern.

A limitation of a more technical nature stems from the fact that the
decision boundaries of the trained models tend to be very sharp, with the
predicted class probabilities being biased towards very close to $0$ or
very close $1$. This has the unfortunate side-effect that the output of a
model cannot also serve as a confidence level of its prediction (which
would be possible if the model did output probabilities around $0.5$).
One solution to this issue, that we consider as a step for future work, 
would be to train an ensemble of models, 
for example 10, and
use the average class probabilities of all the models in the ensemble, 
as the final tremor probability for a subject.

Finally, in its current form, the proposed method operates on
accelerometer signals captured only during phone call events. 
However, the overall
idea is not limited to this data capturing approach and could just as well be
applied to IMU data captured during more general interactions of the user
with their phone, for instance from when the user is typing on the virtual
keyboard of the device. Extending the current approach to work in such cases,
is another direction for future work.

\section{Conclusion}\label{conclusion}
We presented a method for performing binary tremor detection
from accelerometer data obtained in-the-wild, in which
each subject is represented by a bag of acceleration signal segments
and a single tremor label. 
A deep multiple-instance learning approach that combined
feature extraction and a pooling scheme inspired by the attention
mechanism,
was used, in order to identify the key segments within each bag. 
The extensive experiments performed
on a dataset of $45$ subjects, indicate
that the proposed method can indeed identify such instances
and, therefore, successfully handle the
in-the-wild setting of the recorded signals. 
Moreover, our method can be trained efficiently
using only the coarse subject-level annotations available, thus efficiently
handling the problem of weak supervision. Finally, it leads to dramatically improved
performance over the examined alternatives.

\section*{Acknowledgment}
The work leading to these results received funding
from the EU Commission under Grant Agreement No.
690494 (http://www.i-prognosis.eu, H2020).
Data collection was approved by the Institution's
Ethical Review Board and all participating subjects provided electronic
consent.


\bibliographystyle{IEEEtran}
\bibliography{root}








\end{document}